\documentclass[twocolumn,aps,prb]{revtex4}
\usepackage[latin1]{inputenc}
\usepackage{amsmath}
\usepackage{amssymb}
\usepackage{graphicx}

\begin{document}

\title{Angle-dependence of the Hall effect in HgBa$_2$CaCu$_2$O$_6$ thin films}

\author{H.~Richter$^1$, I.~Puica$^1$, W.~Lang$^1$, M.~Peruzzi$^2$,
J.~H.~Durrell$^{2,3}$, H.~Sturm$^2$, J.~D.~Pedarnig$^2$,
D.~B\"{a}uerle$^2$}

\affiliation{$^1$Institut f\"{u}r Materialphysik der
Universit\"{a}t Wien, Boltzmanngasse 5, A-1090 Wien, Austria\\
$^2$Institut f\"{u}r Angewandte Physik,
Johannes-Kepler-Universität Linz, A-4040 Linz, Austria\\
$^3$Department of Materials Science and Metallurgy, University of
Cambridge, Pembroke Street, Cambridge, CB2 3QZ, United Kingdom}

\begin{abstract}
Superconducting compounds of the family Hg-Ba-Ca-Cu-O have been
the subject of intense study since the current record-holder for
the highest critical temperature of a superconductor belongs to
this class of materials. Thin films of the compound with two
adjacent copper-oxide layers and a critical temperature of about
120 K were prepared by a two-step process that consists of the
pulsed-laser deposition of precursor films and the subsequent
annealing in mercury-vapor atmosphere. Like some other
high-temperature superconductors, Hg-Ba-Ca-Cu-O exhibits a
specific anomaly of the Hall effect, a double-sign change of the
Hall coefficient close to the superconducting transition. We have
investigated this phenomenon by measurements of the Hall effect at
different angles between the magnetic field direction and the
crystallographic $c$-axis. The results concerning the upper part
of the transition, where the first sign change occurs, are
discussed in terms of the renormalized fluctuation model for the
Hall conductivity, adapted through the field rescaling procedure
in order to take into account the arbitrary orientation of the
magnetic field.
\end{abstract}


\maketitle

\section{Introduction}

The large anisotropy found in high-temperature superconductors
(HTSC), due to their layered structure, gives rise to important
changes of the transport properties as the magnetic field
orientation varies with respect to the superconductor planes.
Early resistivity and critical current measurements
\cite{Iye89,Amirfeiz97} performed on HTSC as a function of the
angle $\theta$ between the magnetic field and the $ab$-planes have
generally shown that dissipation strongly decreases when the field
is tilted towards the superconducting layers. The measurements at
oblique field orientations were commonly analyzed in terms of the
scaling approach based on the anisotropic mass
model.\cite{Blatter92} According to this scaling approach, the
main effect of the anisotropy is to reduce the field component in
the superconducting planes, such that in the limit of highly
anisotropic materials, the magnetic field component along the
$c$-axis is the only effective one, as is indeed found
experimentally \cite{Iye89,Kes90} on
Bi$_{2}$Sr$_{2}$CaCu$_{2}$O$_{x}$ (BSCCO-2212). For materials with
moderate anisotropy, like YBa$_{2}$Cu$_{3}$O$_{7}$ (YBCO), the
field scaling was confirmed in the flux-flow region
\cite{Harris94} but found however to work not equally well in the
regime of thermal activation of vortices, where the {}``failure of
scaling'' was explained as a consequence of
pinning.\cite{Amirfeiz97} For the compounds of the Hg-Ba-Ca-Cu-O
family, that have an intermediate anisotropy \cite{MSKim96}
between YBCO and BSCCO-2212, many fewer investigations have been
performed regarding the angle-dependence of the transport
properties. The resistivity of (Hg,Re)Ba$_{2}$CaCu$_{2}$O$_{6}$
was recently \cite{Salem04a} studied under variation of the
magnetic field orientation with respect to the $c$-axis, and the
dependence of the depinning field on the tilt angle was inferred.
In this work we present the first investigations of the
Hall-effect's dependence on the angle $\theta$ between the
magnetic field and the $ab$-planes in the
HgBa$_{2}$CaCu$_{2}$O$_{6}$ compound, and compare the experimental
data to theoretical fits based on the renormalized superconducting
fluctuation model.\cite{UD} The model was adapted by a field
rescaling procedure, in order to account for the tilted magnetic
field orientation.

\section{Sample preparation and experimental setup}

Electrical measurements were made on $c$-axis oriented
HgBa$_{2}$CaCu$_{2}$O$_{6}$ (Hg-1212) thin films. The thin films
were fabricated on (001) SrTiO$_3$ crystal substrates in a
two-step process. Amorphous precursor films were deposited on the
substrate using pulsed-laser deposition (PLD) \cite{bauerle98}
and, subsequently, films were annealed in a mercury vapor
atmosphere employing the sealed quartz tube technique.
\cite{yun95,yun96,yun00} Typically, sintered targets of nominal
composition Ba:Ca:Cu = 2:2:3 are employed for laser ablation and
precursor films are deposited at room temperature. For Hg-1212
phase formation and for crystallization of films with $c$-axis
orientation annealing at high temperature ($800-830$~°C) and high
vapor pressure (35 bar) is required. The Hg-Ba-Ca-Cu-O layers
usually reveal reduced surface quality, phase purity and
crystallinity as compared to high-$T_c$ superconducting layers
that are grown in a single-step process (e.g., YBCO and
Bi$_2$Sr$_2$Ca$_{n-1}$Cu$_n$O$_{2(n+2)+\delta}$ films). However,
phase-pure epitaxial Hg-1212 films with improved surface
morphology are achieved by using mercury-doped targets
(Hg:Ba:Ca:Cu $\approx$ 0.8:2:2:3) for laser-ablation and by
deposition of precursor films at higher substrate temperature
($T_S = 350$~°C) \cite{Peruzzi01}.

\begin{figure}
\includegraphics[%
  width=8.5cm]{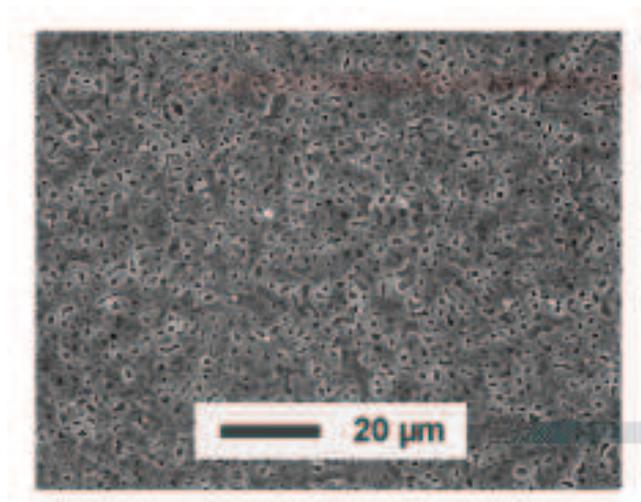}
\caption{\label{struct}Electron microscopy image of a $c$-axis
oriented HgBa$_{2}$CaCu$_{2}$O$_{6}$ thin film on SrTiO$_3$
substrate. Dense and homogeneous layers without irregular surface
structures are fabricated under optimized conditions.}
\end{figure}

Figure~\ref{struct} shows the surface of Hg-1212 layers that were
produced from mercury-doped precursor films deposited at $T_S =
350$~°C. The annealed films reveal a dense and homogeneous
structure and smooth surface without $a$-axis oriented grains and
regions of non-reacted material. The angular spread of the lattice
axes orientation is less than 1° as measured by x-ray diffraction
(rocking curves). For Hall measurements, two Hg-1212 films were
patterned by standard photolithography and wet-chemical etching.
The thickness of films was about 500~nm as measured by atomic
force microscopy and the critical temperature was $T_{c0} \approx
120$~K.

The experimental setup for the electric transport measurements is
made up of a closed-cycle refrigerator and an electromagnet. DC
currents were injected in both directions and the polarity of the
magnetic field was reversed multiple times for every data point to
cancel spurious thermoelectric signals. The longitudinal and
transverse voltages were measured simultaneously. Multiple data
were taken by a Keithley 2182 nanovoltmeter and averaged at
discrete temperature values, with a stability better than $\pm$
0.01 K. The angle $\theta$ between the magnetic field $B$ and the
in-plane current density $j_{x}$ was changed by rotating the
electromagnet for every set of data. The sample was mounted
between the electromagnet's pole pieces in such way that a
standard Hall measurement corresponds to $\theta=90°$. On the
other hand at $\theta=0°$ $B$ is parallel to the current and to
the $ab$-planes of the sample. For all in-field measurements a
magnetic field of $B=1.13$ T was applied. The Hall voltage $V_{y}$
was measured at fixed position on adjacent side arms of the
strip-shaped sample.

\section{results and comparison with theory}

\begin{figure}
\includegraphics[%
  width=8.5cm]{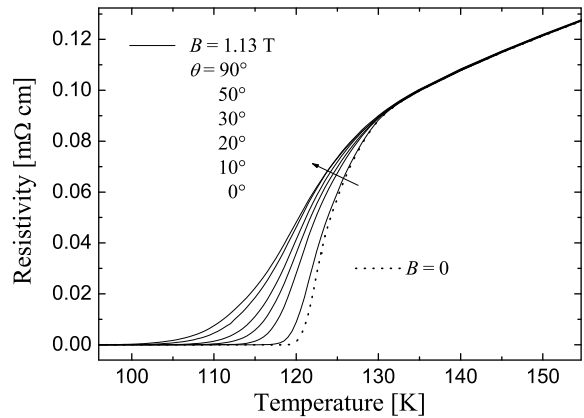}

\caption{\label{RhoXX}Resistivity vs. temperature for different orientations
of a magnetic field of fixed magnitude (solid curves). The arrow indicates
the curve sequence corresponding to the increasing angle $\theta$
of the magnetic field with respect to the $ab$-plane (as well as
to the current direction). The dotted curve shows the zero-field resistivity.}
\end{figure}

Figure \ref{RhoXX} shows the temperature dependence of the
longitudinal resistivity for the sample which had a slightly lower
resistivity, as the magnetic field $B=1.13$~T is rotated in the
plane determined by the $c$-axis and the current direction, at an
angle $\theta$ with the latter. The results for the second sample
are quite similar and are not shown here. The effect of an
increasing $\theta$ at constant magnetic field, namely the
broadening of the superconducting transition and the resistivity
enhancement, is qualitatively similar to that of an increasing
magnetic field applied orthogonally to the layers, as presented
for instance in Ref.~\onlinecite{Hall04}. The difference between
the effect of a perpendicular field $B_{z}$ compared to that of a
tilted one with the same normal component $B_{z} = Bsin\theta$
becomes however evident at small angles, when $B$ is almost
parallel to the layers. As one can see in Fig. \ref{RhoXX}, the
zero-field curve (dotted) is clearly different from the one
corresponding to $B = 1.13$~T at $\theta=0$, where $B_{z} = 0$.
This fact points to the three-dimensional conduction character in
the Hg-Ba-Ca-Cu-O compound, since for a pure two-dimensional
in-plane conduction only the perpendicular-to-surface component is
relevant.\cite{Larkin05}

\begin{figure}
\includegraphics[%
  width=8.5cm]{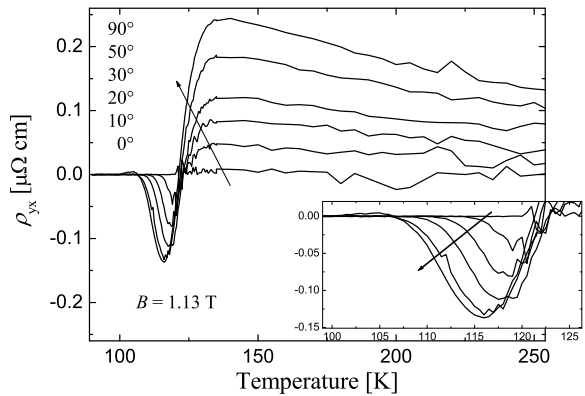}

\caption{\label{RhoHroh}Hall resistivity $\rho_{yx}$ for the same
sample and experimental conditions as in Fig. \ref{RhoXX}. The
arrows indicates the curve sequence corresponding to the increasing
angle $\theta$.}
\end{figure}

The measured Hall resistivity $\rho_{yx} = E_{y}/j_{x}$, where
$E_{y}$ is the electric field between the adjacent side arms of
the sample, is shown in Fig. \ref{RhoHroh}. All the curves, except
that for $\theta=0^{\circ}$, where no Hall signal could be
detected, exhibit the change from the positive, hole-like sign in
the normal state to the negative, electron-like one in the
vortex-liquid regime, in accordance with previous investigations
performed in the similar magnetic field range on other cuprates,
like BSCCO-2223 (Ref.~\onlinecite{Lang95}) and YBCO
(Ref.~\onlinecite{Hall04}). Furthermore, the second sign reversal,
that is the subsequent return of the Hall resistivity to the
positive value before vanishing, is also noticeable for the
perpendicular orientation of the magnetic field (i.e. for
$\theta=90^{\circ}$), as has been also previously observed in this
compound,\cite{Kang2} but is not discernable anymore at oblique
orientations (see the inset in Figs. \ref{RhoHroh}). This fact
agrees with a similar observation on YBCO in Ref.
\onlinecite{Goeb00}, where it was found that the second sign
change disappeared in high current densities or under slightly
tilted field direction, revealing thus its vortex pinning origin.

\begin{figure}
\includegraphics[%
  width=8.5cm]{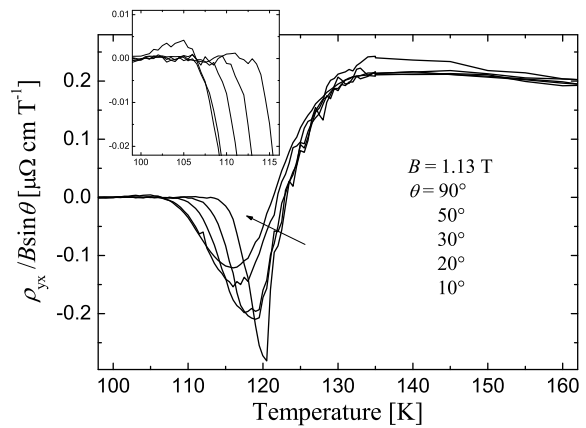}

\caption{\label{RhoH}Hall resistivity $\rho_{yx}$ normalized to the
out-of-plane field component $B\sin\theta$. The arrow shows the
direction of the increasing angle $\theta$. The region of the second
sign change is detailed in the inset.}
\end{figure}

As well as for the longitudinal resistivity, one can notice that,
qualitatively, the Hall resistivity behaves at increasing $\theta$
angle in the same manner as in increasing perpendicular-to-layers
magnetic field $B_{z}$. To explore this situation, the Hall
resistivity normalized to a constant magnetic field component
perpendicular to the $ab$ plane, $B_z$, is shown in Fig.
\ref{RhoH}. Namely, the temperature of the positive-to-negative
sign change shifts to lower values, the negative maximum of the
Hall resistivity normalized to the out-of-plane component of the
applied field becomes less sharp and decreases in magnitude, as
observed for instance also in other cuprates.\cite{Hall04} This
behavior points to the fact that the effect of the magnetic field
is mainly due to its out-of-plane component $B_{z}$.  A good
scaling of the Hall resistivity in the normal state can be
observed, except for $\theta = 10°$, where the large normalization
factor introduces a considerable error.

\begin{figure}
\includegraphics[%
  width=8.5cm]{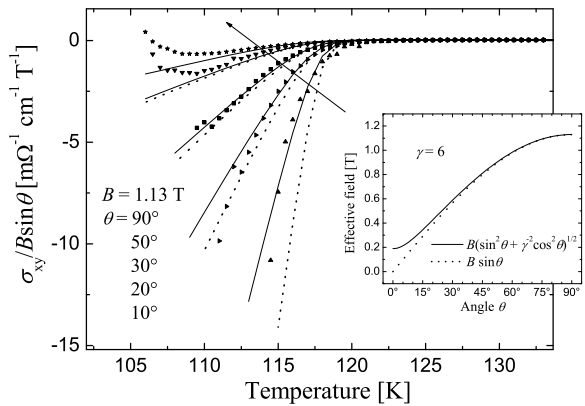}

\caption{\label{SigmaH}Experimental (symbols) and theoretical
(curves) normalized Hall conductivity $\sigma_{xy}/B\sin\theta$ as a
function of temperature. The arrow indicates the data sequence as
$\theta$ increases. The solid curves represent the theoretical fits
based on the rescaled field value $\widetilde{B}$, while for the
dotted curves only the normal component $B\sin\theta$ is considered
relevant. In the inset, the rescaled effective field $\widetilde{B}$
is compared to the out-of-plane field component $B_{z} =
B\sin\theta$.}
\end{figure}

However, more appropriate for a quantitative evaluation and
comparison with a theoretical model for the Hall effect is to
represent the Hall conductivity
$\sigma_{xy}=\rho_{yx}/\left(\rho_{xx}^{2}+\rho_{yx}^{2}\right)$
as a function of temperature, as in Fig. \ref{SigmaH}, where
$\sigma_{xy}$ is shown normalized to $B\sin\theta$. In this
picture, it is easier to discern the delicate interplay of mainly
three contributions to the Hall conductivity:\cite{Hall04}
\textit{(i}) the positive quasiparticle vortex-core contribution,
associated with normal-state excitations; \textit{(ii})
superconducting contribution (excess Hall effect), resulting from
the vortex flux-flow and superconducting fluctuations, which, by
its negative sign, is essential to the occurrence of the first
sign change; and \textit{(iii}) pinning contribution which can
eventually lead to the second sign reversal of the Hall effect.
Theoretical calculations,\cite{Kopnin99} based on a simple model
of pinning potential, suggested that an increasing pinning
strength not only affects the longitudinal flux-flow resistivity,
but also decreased the magnitude of the vortex contribution to the
Hall voltage (fluctuation term in the Ginzburg-Landau approach).
Strong enough pinning can thus result in a second sign reversal of
the Hall resistivity, if the negative vortex (fluctuation)
contribution is reduced in absolute value to magnitudes that are
insufficient to counteract the positive contribution of the normal
state conduction.\cite{Kopnin99,Ikeda99}

We shall try to give a quantitative account for the Hall
conductivity data in Fig. \ref{SigmaH} (symbols) for the temperature
region corresponding to the first sign change and the negative
maximum of $\rho_{yx}$, by considering the normal state contribution
$\sigma_{xy}^{\mathrm{n}}$ and the superconducting fluctuation one
$\Delta\sigma_{xy}$, such as
\begin{equation}
\sigma_{xy}=\sigma_{xy}^{\mathrm{n}}+\Delta\sigma_{xy}\,.
\label{SigmaXY}
\end{equation}
For the normal state part, we assume as valid Anderson's
formula:\cite{Anderson}
$\cot\theta_{H}^{\mathrm{n}}=\sigma_{xx}^{\mathrm{n}}/\sigma_{xy}^{\mathrm{n}}=C_{1}T^{2}+C_{0}$.
We shall assume also a linear temperature dependence of the
longitudinal resistivity in the normal state, so that the
normal-state part of the conductivity tensor consists of the
simple expressions:
\begin{equation}
\sigma_{xx}^{\mathrm{n}}=\frac{1}{p_{0}+p_{1}T}
\:\:\:\text{and}\:\:\:\sigma_{xy}^{\mathrm{n}}=
\frac{1}{p_{0}+p_{1}T}\cdot\frac{1}{C_{1}T^{2}+C_{0}}\:,
\label{NormalState}
\end{equation}
where $p_{0}$, $p_{1}$, $C_{1}$ and $C_{0}$ are to be determined
from fits to the experiment.

The fluctuation Hall conductivity $\Delta\sigma_{xy}$ was first
theoretically calculated by Fukuyama, Ebisawa and Tsuzuki,\cite{FET}
who pointed out that the origin of a non-vanishing Hall current due
to fluctuating Cooper pairs could come from a particle-hole
asymmetry, which reveals a complex relaxation time in the time
dependent Ginzburg-Landau (TDGL) theory. In this early work, it was
implicitly assumed that the fluctuations did not interact; that is,
only Gaussian fluctuation were considered. Accordingly, the
fluctuation parts of the conductivity tensor elements were predicted
to diverge at $T_{c}$ in the presence of magnetic field. However,
this predicted divergence has not been observed. A great improvement
was obtained when the interaction between fluctuations was taken
into account by incorporating the quartic term
$\left|\psi\right|^{4}$ from the Ginzburg-Landau (GL) expression of
the free energy. Such a treatment was performed by Ullah and
Dorsey\cite{UD} (UD) in the frame of a simple Hartree approach of
the TDGL theory, while Nishio and Ebisawa\cite{NE} extended the
calculations of the weak (Gaussian) fluctuation contribution of the
Hall conductivity from Ref. \onlinecite{FET} to the strong
(non-Gaussian) fluctuation regime, based on more sophisticated
renormalization theory by Ikeda, Ohmi and Tsuneto.\cite{IOT} The
renormalized, non-Gaussian fluctuation regime connects therefore the
weak (Gaussian) fluctuation regime in the paraconducting region
above $T_{c}\left(H\right)$ to the vortex liquid (flux-flow) regime
below the mean-field transition, interpolating smoothly without the
$T_{c}$ divergence predicted by the Gaussian theory.

Due to its simpler form, we shall use in this paper the
fluctuation Hall conductivity expression provided by the UD model,
\begin {equation}
\Delta\sigma_{xy}=-\alpha\frac{k_{B}T}{\varepsilon_{F}}
\cdot\frac{e^{2}h^{3}}{2\hbar s}\int_{-\pi}^{\pi}\frac{dq}{2\pi}
\sum_{n=0}^{N_{c}-1}\frac{n+1}{\widetilde{\varepsilon}_{nq}
\widetilde{\varepsilon}_{n+1,q}\widetilde{\varepsilon}_{n+\frac{1}{2},q}^{2}}\,,
\label{DeltaSigmaH}
\end{equation}
where $s$ is the distance between superconducting planes in the
layered model, $h=B/B_{c2}(0)=2\xi_{0}^{2}eB/\hbar$ is the reduced
magnetic field, supposed perpendicular to the layers, and
$\widetilde{\varepsilon}_{nq}=\widetilde{\varepsilon}+
\left(r/2\right)\left(1-\cos q \right)+\left(2n+1\right)h$, with
$r=\left(2\xi_{0c}/s\right)^{2}$, while $\xi_{0}$ and $\xi_{0c}$
are the in-plane and, respectively, out-of-plane coherence lengths
extrapolated at $T=0$. The value of $\alpha$, called the
particle-hole asymmetry parameter,\cite{FET} can be inferred from
the microscopical theory if one considers the energy derivative
$\mathcal{N}'$ of the density of states $\mathcal{N}$ at the Fermi
level $\varepsilon_{F}$, and amounts in the BCS model to
$\alpha=4\varepsilon_{F}\mathcal{N}'/\pi
g_{\mathrm{BCS}}\mathcal{N}^{2}$, with $g_{\mathrm{BCS}}$ ($>0$)
the BCS coupling constant.\cite{FET,NE} The renormalized reduced
temperature $\widetilde{\varepsilon}$ is calculated by solving the
Hartree renormalization equation:\cite{UD}
\begin{eqnarray}
\widetilde{\varepsilon}-\ln\frac{T}{T_{0}} & = & 2gTh
\label{SelfConst}\\
 &  & \cdot\sum_{n=0}^{N_{c}}\frac{1}{\sqrt{\left(\widetilde{\varepsilon}
+h+2nh\right)\left(\widetilde{\varepsilon}+h+2nh+r\right)}}\,,\nonumber
\end{eqnarray}
where $g=2\mu_{0}\kappa_{\mathrm{GL}}^{2}e^{2}\xi_{0}^{2}k_{B}/
\left(\pi\hbar^{2}s\right)$, with $\kappa_{\mathrm{GL}}$ being the
in-plane Ginzburg-Landau parameter
$\kappa_{\mathrm{GL}}=\lambda_{0}/\xi_{0}$, and $T_{0}$ is the
mean-field transition temperature. The relationship between
$T_{0}$ and $T_{c0}$, the actual renormalized critical temperature
in zero-field, will be found by putting
$\widetilde{\varepsilon}=0$ in Eq. (\ref{SelfConst}) after taking
the limit $B\rightarrow0$ and it
writes:\cite{PuicaLangH}
\begin{equation}
T_{0}=T_{c0}\left[\sqrt{c/r}+\sqrt{1+(c/r)}\right]^{2gT_{c0}}\,.
\label{T0-Tc0}
\end{equation}
The sums over the Landau levels (LL) in Eqs. (\ref{DeltaSigmaH})
and (\ref{SelfConst}) must be cut-off at some index $N_{c}$,
reflecting the inherent UV divergence of the Ginzburg-Landau
theory. This procedure suppresses the short wavelength fluctuating
modes through a \emph{momentum} (or, equivalently, \emph{kinetic
energy}) \emph{cut-off} condition, which, in terms of the LL
representation writes \cite{UD,Penev0} $\hbar
eB\left(2n+1\right)\leq c\hbar^{2}/2\xi_{0}^{2}$, with the cut-off
parameter $c$ of the order of unity.

As already mentioned, the above model assumes a magnetic field perpendicular
to the superconducting layers. The more general case with a magnetic
field directed at some arbitrary angle $\theta$ with the $ab$-plane
leads to the appearance of a vector potential in the $c$ direction,
and the problem requires nontrivial calculations.\cite{Larkin05}
In the close vicinity of the transition, however, where the out-of-plane
coherence length grows much larger than the interlayer distance so
that the details of the layered structure are no more relevant, the
anisotropic three-dimensional (3D) fluctuation regime takes place.
In the linear response approximation, i.e. for vanishing electric
fields, even for an arbitrary orientation of the magnetic field, the
transport coefficients in the anisotropic 3D model can be easily treated
by using a special scaling transformation of the coordinates and field
components in the GL equation.\cite{Blatter92,Larkin05} This procedure
reduces the problem to the isotropic case, where the coordinate system
can be in turn freely rotated so that the magnetic field acquires
again only one non-zero component.

Our experimental data correspond to the case when the magnetic field
$B$ is applied in the $xz$-plane, at an angle $\theta$ with the
$x$-axis, which represents the direction of the in-plane current
density $j_{x}$. Following the results from Ref. \onlinecite{Larkin05},
p. 68, we can directly write the Hall conductivity under the tilted
field with the aid of the Hall conductivity under a rescaled field
$\widetilde{B}$ applied along the $c$-direction:
\begin{equation}
\sigma_{xy}\left(\theta,B\right)=\sin\widetilde{\theta}\cdot\sigma_{xy}
\left(0,\widetilde{B}\right)\,,
\label{SigmaXYrescaled}
\end{equation}
where
\begin{eqnarray}
\widetilde{B} & = & B\sqrt{\sin^{2}\theta+\frac{1}{\gamma^{2}}
\cos^{2}\theta}\,,\nonumber \\
\tan\widetilde{\theta} & = & \gamma\tan\theta\,,
\label{B-theta-tilde}\\
\gamma & = & \frac{\xi_{0}}{\xi_{0c}}\,.\nonumber
\end{eqnarray}

Relation (\ref{SigmaXYrescaled}) can be transformed to
\begin{equation}
\frac{\sigma_{xy}\left(\theta,B\right)}{B\sin\theta}=
\frac{\sigma_{xy}\left(0,\widetilde{B}\right)}{\widetilde{B}}\,,
\label{scaledSigmaH}
\end{equation}
which suggests that the experimental data (symbols) in Fig.
\ref{SigmaH} are to be compared with the theoretical results given
by Eqs. (\ref{SigmaXY}-\ref{DeltaSigmaH}) for the normalized Hall
conductivity, provided the rescaled magnitude $\widetilde{B}$ of the
magnetic field is considered. Such calculations are illustrated by
the solid lines in Fig. \ref{SigmaH}, for which the following
parameters, typical for the HgBa$_{2}$CaCu$_{2}$O$_{6}$ compound
were used: $T_{c0}=118.2$ K, $\xi_{0}=1.5$ nm, $\xi_{0c}=0.25$ nm,
$s=1.27$ nm, $\kappa_{\mathrm{GL}}=100$, $\varepsilon_{F}=1000$ K
(in $k_{B}$ units).\cite{MSKim96,Puzniak97} The best fits were
obtained for a particle-hole asymmetry parameter $\alpha=0.26$,
which is comparable with the analogous value found from a similar
fit \cite{Lang95} for BSCCO-2223. The dotted curves in Fig.
\ref{SigmaH} show on the other hand the calculations with the same
parameters, if only the out-of-plane component $B_{z} = B\sin\theta$
were taken into account instead of the rescaled value
$\widetilde{B}$. One can notice that the two approaches give almost
similar results at higher angles, where the rescaled magnitude
$\widetilde{B}$ is almost identical with the normal component
$B\sin\theta$, as illustrated in the inset of Fig. \ref{SigmaH}, but
differ significantly for small angle $\theta$, when the field
direction is close to the $ab$-plane and the difference between
$\widetilde{B}$ and $B\sin\theta$ increases. The good quantitative
agreement between the experimental data and the theoretical fits
performed with the rescaled magnetic field values (solid curves)
points therefore to the fluctuation origin of the Hall effect
behavior in the temperature range corresponding to the first
sign-change and the negative maximum of the Hall resistivity, as
well as to the three-dimensional conduction character in the
Hg-Ba-Ca-Cu-O compound.

\section{Conclusions}

In summary, we have investigated the resistivity of and the Hall
effect in HgBa$_{2}$CaCu$_{2}$O$_{6}$ thin films when the magnetic
field of fixed magnitude is rotated in the plane determined by the
$c$-axis and the electric current direction. At first sight our
measurements indicate qualitative similarity with the behavior in
a variable magnetic field applied perpendicularly to the film,
whose magnitude would be given by the out-of-plane component of
the oblique field. This is also suggested by the rather good
scaling of the Hall resistivity in the normal state with the
out-of-plane magnetic field component (Fig. \ref{RhoH}). However,
the difference between the effect of a perpendicular field
compared to that of a tilted one having the same normal component
is revealed as the magnetic field approaches the orientation
parallel to the layers, and is seen in both the resistivity (Fig.
\ref{RhoXX}) and the Hall conductivity (Fig. \ref{SigmaH})
measurements. The second sign-change of the Hall angle in the
lower part of the transition is found to disappear under oblique
fields, pointing thus to its vortex pinning origin (inset in Fig.
\ref{RhoH}). The good quantitative agreement between the measured
Hall conductivity and the theoretical fits based on the
renormalized superconducting fluctuation model (Fig.
\ref{SigmaH}), adapted through the field rescaling procedure in
order to take into account the arbitrary orientation of the
magnetic field, brings evidence for the anisotropic
three-dimensional conduction character in the Hg-Ba-Ca-Cu-O
compound close to $T_{c}$, as well as for the fluctuation origin
of the Hall effect behavior in the vortex liquid region.

\begin{acknowledgments}
This work was supported by the Austrian Fonds zur F\"{o}rderung der
wissenschaftlichen Forschung and the Micro@Nanofabrication (MNA) Network,
funded by the Austrian Ministry for Economic Affairs and Labour.
\end{acknowledgments}
\bibliographystyle{APSREV}
\bibliography{Herberts}

\end{document}